\documentclass[twocolumn,twocolappendix]{aastex63} 
\usepackage[T5,T1]{fontenc}
\usepackage{subfiles}
\usepackage{aasms} 

\shorttitle{IceCube Search for Neutrino Emission from ULIRGs}
\shortauthors{IceCube Collaboration}

\graphicspath{{./}{figures/}}

\newcommand{\IRAS}{\textit{IRAS}}
\renewcommand{\d}{\mathrm{d}}
\renewcommand{\H}{\mathcal{H}}

\begin{document}

\title{Search for High-Energy Neutrinos from Ultra-Luminous Infrared Galaxies with IceCube}

\affiliation{III. Physikalisches Institut, RWTH Aachen University, D-52057 Aachen, Germany}
\affiliation{Department of Physics, University of Adelaide, Adelaide, 5005, Australia}
\affiliation{Dept. of Physics and Astronomy, University of Alaska Anchorage, 3211 Providence Dr., Anchorage, AK 99508, USA}
\affiliation{Dept. of Physics, University of Texas at Arlington, 502 Yates St., Science Hall Rm 108, Box 19059, Arlington, TX 76019, USA}
\affiliation{CTSPS, Clark-Atlanta University, Atlanta, GA 30314, USA}
\affiliation{School of Physics and Center for Relativistic Astrophysics, Georgia Institute of Technology, Atlanta, GA 30332, USA}
\affiliation{Dept. of Physics, Southern University, Baton Rouge, LA 70813, USA}
\affiliation{Dept. of Physics, University of California, Berkeley, CA 94720, USA}
\affiliation{Lawrence Berkeley National Laboratory, Berkeley, CA 94720, USA}
\affiliation{Institut f{\"u}r Physik, Humboldt-Universit{\"a}t zu Berlin, D-12489 Berlin, Germany}
\affiliation{Fakult{\"a}t f{\"u}r Physik {\&} Astronomie, Ruhr-Universit{\"a}t Bochum, D-44780 Bochum, Germany}
\affiliation{Universit{\'e} Libre de Bruxelles, Science Faculty CP230, B-1050 Brussels, Belgium}
\affiliation{Vrije Universiteit Brussel (VUB), Dienst ELEM, B-1050 Brussels, Belgium}
\affiliation{Department of Physics and Laboratory for Particle Physics and Cosmology, Harvard University, Cambridge, MA 02138, USA}
\affiliation{Dept. of Physics, Massachusetts Institute of Technology, Cambridge, MA 02139, USA}
\affiliation{Dept. of Physics and Institute for Global Prominent Research, Chiba University, Chiba 263-8522, Japan}
\affiliation{Department of Physics, Loyola University Chicago, Chicago, IL 60660, USA}
\affiliation{Dept. of Physics and Astronomy, University of Canterbury, Private Bag 4800, Christchurch, New Zealand}
\affiliation{Dept. of Physics, University of Maryland, College Park, MD 20742, USA}
\affiliation{Dept. of Astronomy, Ohio State University, Columbus, OH 43210, USA}
\affiliation{Dept. of Physics and Center for Cosmology and Astro-Particle Physics, Ohio State University, Columbus, OH 43210, USA}
\affiliation{Niels Bohr Institute, University of Copenhagen, DK-2100 Copenhagen, Denmark}
\affiliation{Dept. of Physics, TU Dortmund University, D-44221 Dortmund, Germany}
\affiliation{Dept. of Physics and Astronomy, Michigan State University, East Lansing, MI 48824, USA}
\affiliation{Dept. of Physics, University of Alberta, Edmonton, Alberta, Canada T6G 2E1}
\affiliation{Erlangen Centre for Astroparticle Physics, Friedrich-Alexander-Universit{\"a}t Erlangen-N{\"u}rnberg, D-91058 Erlangen, Germany}
\affiliation{Physik-department, Technische Universit{\"a}t M{\"u}nchen, D-85748 Garching, Germany}
\affiliation{D{\'e}partement de physique nucl{\'e}aire et corpusculaire, Universit{\'e} de Gen{\`e}ve, CH-1211 Gen{\`e}ve, Switzerland}
\affiliation{Dept. of Physics and Astronomy, University of Gent, B-9000 Gent, Belgium}
\affiliation{Dept. of Physics and Astronomy, University of California, Irvine, CA 92697, USA}
\affiliation{Karlsruhe Institute of Technology, Institute for Astroparticle Physics, D-76021 Karlsruhe, Germany }
\affiliation{Karlsruhe Institute of Technology, Institute of Experimental Particle Physics, D-76021 Karlsruhe, Germany }
\affiliation{Dept. of Physics, Engineering Physics, and Astronomy, Queen's University, Kingston, ON K7L 3N6, Canada}
\affiliation{Dept. of Physics and Astronomy, University of Kansas, Lawrence, KS 66045, USA}
\affiliation{Department of Physics and Astronomy, UCLA, Los Angeles, CA 90095, USA}
\affiliation{Department of Physics, Mercer University, Macon, GA 31207-0001, USA}
\affiliation{Dept. of Astronomy, University of Wisconsin{\textendash}Madison, Madison, WI 53706, USA}
\affiliation{Dept. of Physics and Wisconsin IceCube Particle Astrophysics Center, University of Wisconsin{\textendash}Madison, Madison, WI 53706, USA}
\affiliation{Institute of Physics, University of Mainz, Staudinger Weg 7, D-55099 Mainz, Germany}
\affiliation{Department of Physics, Marquette University, Milwaukee, WI, 53201, USA}
\affiliation{Institut f{\"u}r Kernphysik, Westf{\"a}lische Wilhelms-Universit{\"a}t M{\"u}nster, D-48149 M{\"u}nster, Germany}
\affiliation{Bartol Research Institute and Dept. of Physics and Astronomy, University of Delaware, Newark, DE 19716, USA}
\affiliation{Dept. of Physics, Yale University, New Haven, CT 06520, USA}
\affiliation{Dept. of Physics, University of Oxford, Parks Road, Oxford OX1 3PU, UK}
\affiliation{Dept. of Physics, Drexel University, 3141 Chestnut Street, Philadelphia, PA 19104, USA}
\affiliation{Physics Department, South Dakota School of Mines and Technology, Rapid City, SD 57701, USA}
\affiliation{Dept. of Physics, University of Wisconsin, River Falls, WI 54022, USA}
\affiliation{Dept. of Physics and Astronomy, University of Rochester, Rochester, NY 14627, USA}
\affiliation{Department of Physics and Astronomy, University of Utah, Salt Lake City, UT 84112, USA}
\affiliation{Oskar Klein Centre and Dept. of Physics, Stockholm University, SE-10691 Stockholm, Sweden}
\affiliation{Dept. of Physics and Astronomy, Stony Brook University, Stony Brook, NY 11794-3800, USA}
\affiliation{Dept. of Physics, Sungkyunkwan University, Suwon 16419, Korea}
\affiliation{Institute of Basic Science, Sungkyunkwan University, Suwon 16419, Korea}
\affiliation{Dept. of Physics and Astronomy, University of Alabama, Tuscaloosa, AL 35487, USA}
\affiliation{Dept. of Astronomy and Astrophysics, Pennsylvania State University, University Park, PA 16802, USA}
\affiliation{Dept. of Physics, Pennsylvania State University, University Park, PA 16802, USA}
\affiliation{Dept. of Physics and Astronomy, Uppsala University, Box 516, S-75120 Uppsala, Sweden}
\affiliation{Dept. of Physics, University of Wuppertal, D-42119 Wuppertal, Germany}
\affiliation{DESY, D-15738 Zeuthen, Germany}

\author[0000-0001-6141-4205]{R. Abbasi}
\affiliation{Department of Physics, Loyola University Chicago, Chicago, IL 60660, USA}

\author[0000-0001-8952-588X]{M. Ackermann}
\affiliation{DESY, D-15738 Zeuthen, Germany}

\author{J. Adams}
\affiliation{Dept. of Physics and Astronomy, University of Canterbury, Private Bag 4800, Christchurch, New Zealand}

\author[0000-0003-2252-9514]{J. A. Aguilar}
\affiliation{Universit{\'e} Libre de Bruxelles, Science Faculty CP230, B-1050 Brussels, Belgium}

\author[0000-0003-0709-5631]{M. Ahlers}
\affiliation{Niels Bohr Institute, University of Copenhagen, DK-2100 Copenhagen, Denmark}

\author{M. Ahrens}
\affiliation{Oskar Klein Centre and Dept. of Physics, Stockholm University, SE-10691 Stockholm, Sweden}

\author{C. Alispach}
\affiliation{D{\'e}partement de physique nucl{\'e}aire et corpusculaire, Universit{\'e} de Gen{\`e}ve, CH-1211 Gen{\`e}ve, Switzerland}

\author{A. A. Alves Jr.}
\affiliation{Karlsruhe Institute of Technology, Institute for Astroparticle Physics, D-76021 Karlsruhe, Germany }

\author{N. M. Amin}
\affiliation{Bartol Research Institute and Dept. of Physics and Astronomy, University of Delaware, Newark, DE 19716, USA}

\author{R. An}
\affiliation{Department of Physics and Laboratory for Particle Physics and Cosmology, Harvard University, Cambridge, MA 02138, USA}

% ...

\collaboration{381}{IceCube Collaboration}

\submitjournal{ApJ}
\accepted{November 22, 2021}

\begin{abstract}
Ultra-luminous infrared galaxies (ULIRGs) have infrared luminosities $L_{\mathrm{IR}} \geq 10^{12} L_{\odot}$, making them the most luminous objects in the infrared sky. These dusty objects are generally powered by starbursts with star-formation rates that exceed $100~ M_{\odot}~ \mathrm{yr}^{-1}$, possibly combined with a contribution from an active galactic nucleus. Such environments make ULIRGs plausible sources of astrophysical high-energy neutrinos, which can be observed by the IceCube Neutrino Observatory at the South Pole. We present a stacking search for high-energy neutrinos from a representative sample of 75 ULIRGs with redshift $z \leq 0.13$ using 7.5 years of IceCube data. The results are consistent with a background-only observation, yielding upper limits on the neutrino flux from these 75 ULIRGs. For an unbroken $E^{-2.5}$ power-law spectrum, we report an upper limit on the stacked flux $\Phi_{\nu_\mu + \bar{\nu}_\mu}^{90\%} = 3.24 \times 10^{-14}~ \mathrm{TeV^{-1}~ cm^{-2}~ s^{-1}}~ (E/10~ \mathrm{TeV})^{-2.5}$ at 90\% confidence level. In addition, we constrain the contribution of the ULIRG source population to the observed diffuse astrophysical neutrino flux as well as model predictions.
\end{abstract}

\section{Introduction}
\label{sec:intro}

The observation of high-energy astrophysical neutrinos with IceCube \citep{IceCube_discovery_science,IceCube_discovery_PRD} marked the birth of neutrino astronomy. Numerous studies have been performed searching for the sources of these astrophysical neutrinos, which are cosmic messengers that represent a smoking-gun signature of hadronic acceleration. So far, the blazar TXS 0506+056 \citep{IceCube_TXS_170922a,IceCube_TXS_flare} is the sole neutrino source candidate that has been identified with a significance at the $3\sigma$ level, and the first indications of neutrino emission have been found from the starburst galaxy NGC 1068 \citep{IceCube_PS_10yr}. Studies from the community have also shown indications of astrophysical neutrinos correlated with the tidal disruption event AT2019dsg \citep{Stein_2021} and radio-bright active galactic nuclei \citep[AGN;][]{Plavin_2020}. However, these results are insufficient to explain the origin of the observed diffuse astrophysical neutrino flux \citep{IceCube_diffuse_combined,IceCube_diffuse_numu_6yr,IceCube_diffuse_cascades_6yr,IceCube_diffuse_HESE_7.5yr}. \added{Further dedicated IceCube studies (see \citealt{IceCube_ICRC2021} for an overview of the latest IceCube searches) have investigated a wide range of candidate neutrino-source classes and found no evidence for neutrinos originating from such sources.} Constraints from \added{these} IceCube studies imply that if the neutrino sky is dominated by a single source population, this population likely consists of relatively numerous, low-luminosity sources \citep{Lipari_2008,Silvestri_2010,Murase_2012,Ahlers_2014,Kowalski_2015,Murase_2016,IceCube_numu_ps_8yr,IceCube_transient_constraints,Ackermann_2019}. In particular, the study of \citet{IceCube_2LAC_constraints} constrains the diffuse neutrino contribution of blazars in the 2LAC catalog of the \textit{Fermi} Large Area Telescope \citep[LAT;][]{Fermi_LAT,Fermi_2LAC}. Furthermore, \textit{Fermi}-LAT observations provide an upper bound on the contribution of non-blazar sources to the extragalactic gamma-ray background \citep[EGB;][]{Fermi_IGRB,Fermi_EGB}. This bound implies constraints on the non-blazar source populations that could be responsible for the diffuse neutrino flux \citep[e.g.][]{Bechtol_2017}. In particular, the non-blazar EGB bound hints towards neutrino sources that are gamma-ray opaque \citep{Murase_2016_hidden_sources,Vereecken_2020}. 


In this work we investigate ultra-luminous infrared galaxies (ULIRGs; see \citealt{Lonsdale_2006} for a review) as neutrino source candidates. These objects are characterized by a rest-frame infrared (IR) luminosity\footnote{Note that in this work we do not distinguish hyper-luminous infrared galaxies (HyLIRGs; $L_{\mathrm{IR}} \geq 10^{13} L_{\odot}$) from the ULIRG source class.} $L_{\mathrm{IR}} \geq 10^{12} L_{\odot}$ between 8--1000 $\mu$m. ULIRG morphologies typically contain features of spiral-galaxy mergers \citep[e.g.][]{Hung_2014}, indicating that they correspond to an evolutionary phase of such systems \citep[e.g.][]{Larson_2016}. The local source density of ULIRGs is ${\sim}10^{-7}\text{--}10^{-6}~ \mathrm{Mpc^{-3}}$, although the abundance of ULIRGs increases rapidly up to a redshift $z \sim 1$ \citep{Kim_1998a,Blain_1999,Cowie_2004,LeFloch_2005,HopkinsPF_2006,Magnelli_2009,Clements_2010,Goto_2011,Magnelli_2011,Casey_2012}. ULIRGs and their less luminous but more numerous counterparts, the luminous infrared galaxies (LIRGs; $10^{11} L_{\odot} \leq L_{\mathrm{IR}} < 10^{12} L_{\odot}$), become the main contributors to the IR energy density at redshifts $z \sim 2\text{--}3$ \citep{Genzel_2000,Chapman_2005,Reddy_2008}.

ULIRGs are predominantly powered by starbursts with star-formation rates $\gtrsim 100~ M_{\odot}~ \mathrm{yr^{-1}}$ \citep{Lonsdale_2006,Rieke_2009,daCunha_2010,Lopez_2016}. The strong thermal IR emission of ULIRGs is the result of abundant dust and gas reprocessing higher-frequency radiation. In addition,  can show signs of AGN activity \citep{Nagar_2003,Lonsdale_2006,Clements_2010,Fadda_2010}. The fraction of ULIRGs containing an AGN is $\gtrsim 40\%$, with higher values reported as a function of increasing infrared luminosity \citep{Veilleux_1995,Kim:1998b,Veilleux_1999,Goto_2005,Imanishi_2008,Hou_2009}. The AGN contribution to the infrared luminosity is typically of the order of $\sim$10\%, although AGN can become the dominant contributors for ULIRGs with a stronger infrared output \citep{Farrah_2003,Veilleux_2009,Imanishi_2010,Nardini_2010,Yuan_2010}. \citet{Lonsdale_2006} also note that the AGN power could be underestimated due to strong obscuration of the AGN. This result is consistent with the work of \citet{Nardini_2011}, who find that AGN in a sample of local ULIRGs are Compton-thick, i.e.~the AGN are obscured by dust clouds with column densities $N_H \gtrsim 1.5 \times 10^{24}~ \mathrm{cm^{-2}}$ \citep{Comastri_2004}. Observations of Arp 220, i.e.~the closest ULIRG, indicate that a possible AGN hosted by this object should be Compton-thick with $N_H \gtrsim 10^{25}~ \mathrm{cm^{-2}}$ \citep{Wilson_2014,Scoville_2017}.

Hadronic acceleration, and hence neutrino production, could occur both in starburst reservoirs \citep[e.g.][]{Tamborra_2014,Peretti_2019,Ambrosone_2021} and in AGN \citep[e.g.][]{Murase_2017,Inoue_2019,Kheirandish_2021}. Since ULIRGs host such environments, this implies that ULIRGs are candidate sources of astrophysical neutrinos. \citet{He_2013} have constructed a reservoir model of the starburst regions of ULIRGs, where the enhanced hypernova rates are responsible for hadronic acceleration. They predict a PeV diffuse neutrino flux from ULIRGs that can explain a significant fraction of the diffuse IceCube observations. The generic reservoir model of \citet{Palladino_2019} considers a neutrino flux from hadronically-powered gamma-ray galaxies (HAGS), which include ULIRGs and starburst galaxies with $L_{\mathrm{IR}} < 10^{12} L_{\odot}$. For power-law spectra ($\Phi_{\nu + \bar{\nu}} \propto E^{-\gamma}$) with spectral indices $\gamma \leq 2.12$, this model can partially explain the diffuse neutrino observations while remaining consistent with the non-blazar EGB bound above 50 GeV \citep{Fermi_EGB,Lisanti_2016,Zechlin_2016}. In particular, a population of HAGS could be the dominant contributor to the diffuse neutrino flux above 100 TeV, and it could be responsible for roughly half of the diffuse neutrino flux between 10--100 TeV. In contrast with such reservoir scenarios, \citet{Vereecken_2020} model the neutrino production in ULIRGs through a beam dump of hadrons accelerated in a Compton-thick AGN. They find that if the AGN is obscured by column densities $N_H \gtrsim 5 \times 10^{25}~ \mathrm{cm^{-2}}$, ULIRGs could fit the diffuse neutrino observations without violating the \textit{Fermi}-LAT non-blazar EGB bound above 50 GeV \citep{Fermi_EGB}.

This paper presents an IceCube search for high-energy astrophysical neutrinos originating from ULIRGs. In Section \ref{sec:ULIRG_search} we describe the source selection, the IceCube dataset, and the analysis method used in this search. The results of the analysis are presented in Section \ref{sec:results}. The interpretation of these results is given in Section \ref{sec:discussion}, where we discuss the implications for the neutrino flux from the ULIRG source population, and where we compare our results to the models of \citet{He_2013}, \citet{Palladino_2019}, and \citet{Vereecken_2020}. Finally, we present our conclusions in Section \ref{sec:conclusions}. In this work, we assume a $\Lambda$CDM cosmology using the \textit{Planck} 2015 results \citep{Planck_2015}, $H_0 = 67.8~\mathrm{km~ s^{-1}~ Mpc^{-1}}$, $\Omega_m = 0.31$, and $\Omega_{\Lambda} = 0.69$.

\section{Search for Neutrino Emission from ULIRGs}
\label{sec:ULIRG_search}

\subsection{Selection of ULIRGs}
\label{sec:ULIRG_selection}

The ULIRGs for this analysis are obtained from three different catalogs, primarily based on data from the Infrared Astronomical Satellite \citep[\IRAS;][]{Neugebauer_1984}, listed below. The infrared luminosity $L_{\mathrm{IR}}$ listed in the catalogs is determined using the \IRAS~flux measurements at 12 $\mu$m, 25 $\mu$m, 60 $\mu$m, and 100 $\mu$m \citep[see e.g.][]{Sanders_1996}. The three catalogs are:
\begin{enumerate}
    \item The \IRAS~Revised Bright Galaxy Sample \citep[RBGS;][]{Sanders_2003}. This catalog contains the brightest extragalactic sources observed by \IRAS. These RBGS objects are sources with a 60 $\mu$m infrared flux $f_{60} > 5.24~ \mathrm{Jy}$ and a Galactic latitude $\left| b \right| > 5^{\circ}$ to exclude the Galactic Plane. The RBGS provides the total infrared luminosity between 8--1000 $\mu$m for all objects in the sample, containing 21 ULIRGs.
    
    \item The \IRAS~1 Jy Survey of ULIRGs \citep{Kim_1998a} selected from the \IRAS~Faint Source Catalog \citep[FSC;][]{Moshir_1992} contains sources with $f_{60} > 1~ \mathrm{Jy}$. The survey required the ULIRGs to have a Galactic latitude $\left| b \right| > 30^{\circ}$ to avoid strong contamination from the Galactic Plane. Furthermore, in order to have accessible redshift information from observatories located at Mauna Kea, Hawaii, this survey is restricted to declinations $\delta > -40^{\circ}$. The resulting selection is a set of 118 ULIRGs.
    
    \item The ULIRG sample used by \citet{Nardini_2010}. Their selection is primarily based on the redshift survey \citep{Saunders_2000} of the \IRAS~Point Source Catalog \citep[PSC;][]{Beichman_1988}. This catalog contains objects with $f_{60} \gtrsim 0.6~ \mathrm{Jy}$ and covers 84\% of the sky. In addition, the authors require the ULIRGs to be observed by the Infrared Spectograph \citep[IRS;][]{Houck_2004} on the \textit{Spitzer Space Telescope} \citep{Werner_2004}. As such, they obtain a sample of 164 ULIRGs.
\end{enumerate}

There exists some overlap between the ULIRGs of the above three catalogs. Therefore, the NASA/IPAC Extragalactic Database\footnote{Available at \href{https://ned.ipac.caltech.edu/}{ned.ipac.caltech.edu}.} (NED) is used to cross-identify these sources. This cross-identification was done by obtaining the NED source list of each catalog, and then cross-identifying sources with the same NED information in each list. The result is a selection of 189 unique ULIRGs. For uniformity, NED is also used to obtain the equatorial coordinates and redshift for each object. Since $L_{\mathrm{IR}}$ values depend on \IRAS~flux measurements, which are optimized separately for the different \IRAS~surveys, they are taken in the following order:
\begin{itemize}
    \item From Catalog 1 if available;
    \item From Catalog 2 if not available in Catalog 1;
    \item From Catalog 3 if not available in Catalog 1 or 2.
\end{itemize}
Note that Catalog 3 is the only catalog that provides uncertainties on these values. We therefore include all objects from this catalog that are consistent with $L_{\mathrm{IR}} = 10^{12} L_{\odot}$ within one standard deviation\footnote{The NED identifications (with $\log_{10}(L_{\mathrm{IR}} / L_{\odot})$ values) of these objects are UGC 05101 ($11.99 \pm 0.02$), IRAS 18588+3517 ($11.97 \pm 0.04$), and 2MASX J23042114+3421477 ($11.99 \pm 0.04$).}.

\begin{figure}
    \centering
    \includegraphics[width=0.9\linewidth]{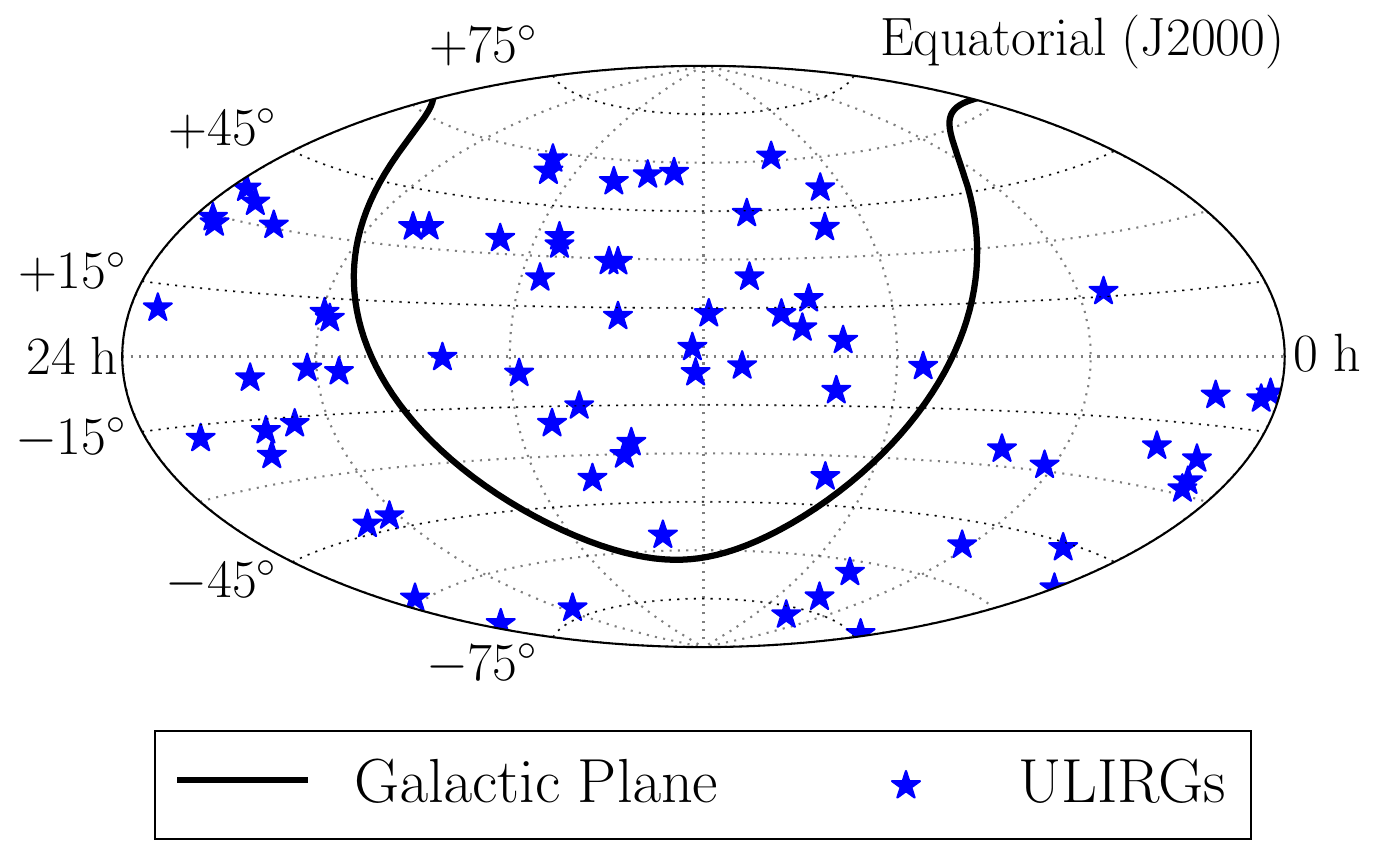}
    \caption{Sky distribution of the representative sample of 75 ULIRGs with redshift $z \leq 0.13$ selected for this IceCube search.}
    \label{fig:skymap_ULIRGs}
\end{figure}

To obtain a representative sample of the local ULIRG population, which we define as a sample of ULIRGs that is complete up to a given redshift, we place a cut on the redshifts in our initial ULIRG selection. We find that keeping only those objects with $z \leq 0.13$ allows the least luminous ULIRGs (i.e.~$L_{\mathrm{IR}} = 10^{12} L_{\odot}$) to be observed, given a conservative \IRAS~sensitivity $f_{60} = 1~ \mathrm{Jy}$ (see Appendix \ref{app:completeness}, where we also take into account the effect of the limited sky coverage of Catalog 2). The result is a representative sample of 75 ULIRGs with $z \leq 0.13$, shown in Fig.~\ref{fig:skymap_ULIRGs}. Table \ref{tab:ulirgs} lists their NED identification, equatorial coordinates, redshifts, fluxes at 60 $\mu$m, and total IR luminosities.

\subsection{Detector and Data Set}

The IceCube Neutrino Observatory is a $1~ \mathrm{km}^3$ optical Cherenkov detector located at the geographic South Pole \citep{IceCube_detector}. The detector is buried in the ice at depths between 1450 m and 2450 m below the surface. It consists of 5160 digital optical modules (DOMs), which are distributed over 86 vertical strings. Each DOM contains a photomultiplier tube and on-board read-out electronics \citep{IceCube_DAQ,IceCube_DOM}. When a high-energy neutrino interacts with the ice or the bedrock in the vicinity of the detector, secondary relativistic particles are produced that emit Cherenkov radiation, which can be registered by the DOMs. The number of photons collected by the DOMs gives a measure of the energy deposited by these secondary charged particles. This feature combined with the arrival time of the light at the DOMs is used to reconstruct the particle trajectories, which are subsequently used to determine the arrival direction of the original neutrino.

IceCube is sensitive to all neutrino flavors, although neutrinos and antineutrinos can generally not be distinguished. Neutral-current interactions of the three flavors, and charged-current interactions of electron neutrinos ($\nu_e + \bar{\nu}_e$) and tau neutrinos ($\nu_\tau + \bar{\nu}_\tau$) are observed by means of their particle cascades. These cascades lead to a rather spherical pattern of recorded photons inside the detector, with a relatively poor angular resolution \citep[$\gtrsim 8^\circ$; see][]{IceCube_cascades}. However, charged-current interactions of muon neutrinos ($\nu_\mu + \bar{\nu}_\mu$) produce muons, which can leave extended track-like signatures in the detector\footnote{Tracks can also be the signature of taus produced in charged-current interactions of tau neutrinos. However, such tracks only become detectable at the highest tau energies $E_{\tau}$, since the decay length of the tau is roughly $\mathrm{50~ m} \times (E_{\tau}/\mathrm{PeV})$ \citep[see also][]{IceCube_tau}.}. The typical angular resolution of these tracks is $\lesssim 1^{\circ}$ for muons with energies $\gtrsim 1~ \mathrm{TeV}$ \citep{IceCube_realtime_alert,IceCube_PS_10yr}.

For this analysis we use the IceCube gamma-ray follow-up (GFU) data sample \citep{IceCube_realtime_alert}, which contains well-reconstructed muon tracks with $4\pi$ sky coverage. The track selection is performed separately for both hemispheres due to their differing background characteristics. The background for astrophysical neutrinos originates from cosmic-ray induced particle cascades in Earth's atmosphere. Atmospheric neutrinos produced in such cascades are able to reach the detector from both hemispheres, leading to a background event rate at the mHz level. Compared to astrophysical neutrinos, atmospheric neutrinos have a relatively soft spectrum that dominates at energies $\lesssim 100~ \mathrm{TeV}$ \citep{IceCube_atmospheric,IceCube_diffuse_combined,IceCube_diffuse_numu_6yr,IceCube_diffuse_cascades_6yr,IceCube_diffuse_HESE_7.5yr}. Atmospheric muons produced in the Northern hemisphere are attenuated by Earth before reaching IceCube. In the Southern hemisphere, however, they are able to penetrate the Antarctic ice and leave track signatures in the detector. These atmospheric muons have a median event rate of 2.7 kHz, requiring a more stringent event selection in the Southern sky. The event selection criteria result in a GFU track sample at the atmospheric neutrino level, with an all-sky event rate of 6.6 mHz. Note that this is still various orders of magnitude above the expected rate of astrophysical neutrinos at the $\mu$Hz level. The selection efficiency of the GFU sample is determined by simulating signal neutrinos according to an unbroken power-law spectrum $\Phi_{\nu + \bar{\nu}} \propto E_{\nu}^{-2}$ between $100~ \mathrm{GeV} < E_{\nu} < 1~ \mathrm{EeV}$. In the Northern hemisphere, the selection efficiency is $\sim$50\% at $E_{\nu} \sim 100~ \mathrm{GeV}$ and reaches $\sim$95\% for $E_{\nu} \gtrsim 100~ \mathrm{TeV}$. In the Southern hemisphere, the selection efficiency is $\sim$5\% at $E_{\nu} \sim 10~ \mathrm{TeV}$ and exceeds 70\% for $E_{\nu} \gtrsim 1~ \mathrm{PeV}$. The GFU data used in this search contains over $1.5 \times 10^6$ events, spanning a livetime of 2615.97 days $\sim$ 7.5 years of the full 86-string IceCube configuration between 2011--2018.

\subsection{Analysis Method}
\label{sec:analysis_method}

In this work, we search for an astrophysical component in the IceCube data which is spatially correlated with our representative sample of 75 ULIRGs. This astrophysical signal would be characterized by an excess of neutrino events above the background of atmospheric muons and atmospheric neutrinos. To search for this excess, a time-integrated unbinned maximum likelihood analysis is performed \citep{Braun_2008}. In addition, the ULIRGs are stacked in order to enhance the sensitivity of the analysis \citep[see][]{IceCube_stacking}.

The unbinned likelihood for a search stacking $M$ candidate point sources using $N$ data events is given by
\begin{equation}
    \mathcal{L}(n_s,\gamma) = \prod_{i=1}^{N} \left[ \frac{n_s}{N} \sum_{k=1}^{M} w_k \mathcal{S}_i^k(\gamma) + \left( 1 - \frac{n_s}{N} \right) \mathcal{B}_i \right].
\label{eq:llh}
\end{equation}
Here, $n_s$ denotes the number of signal events, and $\gamma$ denotes the spectral index of the signal energy distribution, which is assumed to follow an unbroken power-law spectrum, $\Phi_{\nu + \bar{\nu}} \propto E^{-\gamma}$. These fit parameters are restricted to $n_s \geq 0$ and $\gamma \in [1,4]$ but allowed to float otherwise. For each event $i$, the probability density functions (PDFs) of the background, $\mathcal{B}_i$, and the signal for source $k$, $\mathcal{S}_i^k$, are evaluated\added{\footnote{In the unbinned likelihood formulation of Eq.~(\ref{eq:llh}), the PDFs are assumed to be continuous \citep{Braun_2008}. This continuity is approximated by choosing a PDF bin size that is well below the angular resolution ($\lesssim 1^\circ$) and energy resolution ($\log(E/\mathrm{GeV}) \sim 0.3$) for track-like events.}}. The parameter $w_k$ represents the stacking weight of source $k$.

Both the signal PDF and background PDF are separated into spatial and energy components. The background PDF is evaluated as $\mathcal{B}_i = B(\delta_i)/2\pi \times \mathcal{E}_{\mathcal{B}}(E_i)$, with $\delta_i$ and $E_i$ the reconstructed declination and energy of event $i$, respectively. The spatial component is uniform in right ascension due to the rotation of the IceCube detector with Earth's axis, resulting in a factor $1/2\pi$. The spatial PDF, $B$, and the energy PDF, $\mathcal{E}_{\mathcal{B}}$, are constructed from experimentally obtained background data in reconstructed declination and energy. Analogously, the signal PDF of candidate point source $k$ is evaluated as $\mathcal{S}_i^k(\gamma) = S(\textbf{x}_{k},\textbf{x}_{i},\sigma_i) \times \mathcal{E}_{\mathcal{S}}(E_i,\gamma)$. For the spatial PDF of the signal, $S$, we take a two-dimensional Gaussian centered around the location of the source, $\textbf{x}_{k}$. The Gaussian is evaluated using the reconstructed location of event $i$, $\textbf{x}_{i}$, and its angular uncertainty, $\sigma_i$, as the standard deviation. The estimation of the angular uncertainty does not take into account systematic effects, such that we follow \citet{IceCube_PS_10yr} and impose a lower bound $\sigma_i \geq 0.2^{\circ}$. Since $\sigma_i$ is much larger than the uncertainty on the source location, the latter may safely be neglected. \replaced{The energy PDF of the signal, $\mathcal{E}_{\mathcal{S}}$, is modeled using an unbroken power-law spectrum with spectral index $\gamma \in [1,4]$. We assume that all candidate sources have identical accelerator properties, and we therefore use a single spectral index $\gamma$ to represent the neutrino spectrum of all ULIRGs.}{The energy PDF of the signal, $\mathcal{E}_{\mathcal{S}}$, is modeled using a power-law spectrum with spectral index $\gamma \in [1,4]$, where we fit a single value of $\gamma$ for the stacked flux of all ULIRGs. This modeling is motivated by the diffuse neutrino observations, which are currently best described by a power law \citep[e.g.][]{IceCube_diffuse_HESE_7.5yr,IceCube_diffuse_numu_9.5yr}.}



The stacking weight of source $k \in \lbrace 1,2,...,M \rbrace$ is constructed as $w_k = r_k t_k / \sum_{j=1}^{M} r_j t_j$. Firstly, it depends on the detector response for an unbroken $E^{-\gamma}$ power-law spectrum at the source declination $\delta_k$. This dependence is reflected by the weight term $r_k \propto \int A_{\mathrm{eff}} (E,\delta_k) E^{-\gamma} \d E$, where $A_{\mathrm{eff}}$ is the effective area of the detector \citep{IceCube_realtime_alert}. Secondly, the stacking weight depends on a theoretical weight $t_k$, which is modeled according to the hypothesis tested by the analysis. In this work, we make the generic assumption that the total IR luminosity between 8--1000 $\mu$m, $L_{\mathrm{IR}}$, is representative for the power of a possible hadronic accelerator in ULIRGs. This assumption is motivated by the fact that the total IR luminosity is directly related to the star-formation rate \citep[see e.g.][]{Rieke_2009}, and that a possible AGN contribution likely increases with IR luminosity \citep[see e.g.][]{Nardini_2010}. Additionally, we assume that the neutrino production is identical in all candidate sources. Hence, we set $t_k = F_{\mathrm{IR},k}$, which is the total IR flux (8--1000 $\mu$m) of ULIRG $k$. The total infrared flux is computed as $F_{\mathrm{IR}} = L_{\mathrm{IR}}/(4 \pi d_L^2)$, where $L_{\mathrm{IR}}$ is taken from the respective ULIRG catalog (see Section \ref{sec:ULIRG_selection}), and where $d_L$ is the luminosity distance determined from the redshift measurements. \added{We note that the theoretical weight is predominantly driven by the luminosity distance, since the IR luminosity values span less than an order of magnitude.}


Subsequently, a test statistic is constructed as $\mathrm{TS} = 2 \log\mathcal{L}(n_s=\hat{n}_s,\gamma=\hat{\gamma}) / \log\mathcal{L}(n_s=0)$. Here, $\hat{n}_s$ and $\hat{\gamma}$ are those values that maximize the likelihood given in Eq.~(\ref{eq:llh}), where a single $\hat{\gamma}$ is fitted for the full ULIRG sample. This TS is used to differentiate data containing a signal component from data being compatible with background. The background-only TS distribution is determined by randomizing the \replaced{IceCube}{GFU} data in right ascension $10^5$ times, and calculating the TS in each iteration, which is called a trial. However, ${\sim}10^8$ trials are required to obtain a background-only TS distribution that is accurate up to the $5\sigma$ significance level. As this is computationally unfeasible, we \replaced{use a $\chi^2$ PDF to approximate the background-only distribution \citep{Wilks_1938}}{fit a $\chi^2$ PDF to the $10^5$ background-only trials in order to approximate the background-only TS distribution \citep{Wilks_1938}}. This PDF is then used to compute the one-sided p-value corresponding with a certain observed $\mathrm{TS_{obs}}$, \added{which is the unique TS value associated with the unscrambled GFU data.} The p-value reflects the compatibility of \deleted{a certain} $\mathrm{TS_{obs}}$ with the background-only scenario. This p-value is determined by evaluating the background-only survival function at $\mathrm{TS_{obs}}$, i.e.~the integral of the background-only PDF over all $\mathrm{TS \geq TS_{obs}}$.

\begin{figure*}
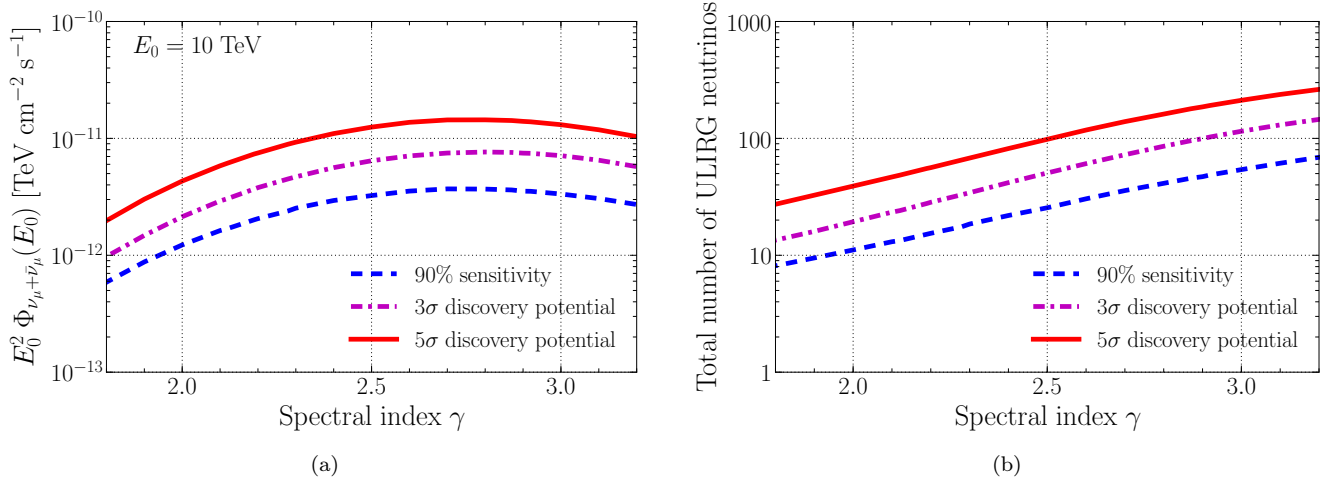

    \centering
    \gridline{\fig{{ULIRG_sensitivity_flux}.pdf}{0.47\textwidth}{(a)}
              \fig{{ULIRG_sensitivity_Nevents}.pdf}{0.47\textwidth}{(b)}}
    \caption{Sensitivity, $3\sigma$, and $5\sigma$ discovery potentials as a function of the spectral index $\gamma$, for injected pseudo-signal following an unbroken $E^{-\gamma}$ power-law spectrum. Panel (a) shows these quantities in terms of the flux at the normalization energy $E_0 = 10~ \mathrm{TeV}$. Panel (b) shows them in terms of the mean number of injected pseudo-signal events. This number is found by convolving the injection spectrum with the effective area of the detector, and integrating it over energy and detector livetime.}
    \label{fig:sensitivity}
\end{figure*}

To test the performance of the analysis, we simulate muon neutrinos with a true direction exactly at the location of our 75 selected ULIRGs. The relative number of neutrinos simulated from an ULIRG location is proportional to the stacking weight of that source. These neutrinos, with energies $100~ \mathrm{GeV} < E_{\nu} < 100~ \mathrm{PeV}$, are generated from an unbroken power-law spectrum, $\Phi_{\nu_\mu + \bar{\nu}_\mu}(E_\nu) = \Phi_{\nu_\mu + \bar{\nu}_\mu}(E_0) (E_\nu/E_0)^{-\gamma}$, which is normalized at an energy\footnote{Note that for all spectra considered in this work, the analysis is sensitive to neutrinos with an energy $E_0 = 10~ \mathrm{TeV}$.} $E_0 = 10~ \mathrm{TeV}$. \added{Here we assume that all sources have identical accelerator properties, such that we use a single spectral index $\gamma$ to represent the flux of all ULIRGs.} In such a simulation, the neutrino produces a track signature in the detector, i.e.~a pseudo-signal event. The angle between the reconstructed direction of the track and the true direction of the neutrino creates a spread of these pseudo-signal events centered around the ULIRG locations. The number of pseudo-signal events are drawn from a Poisson distribution with a given mean. For various values of this mean pseudo-signal, we determine the corresponding TS distribution by performing $10^3$ trials. We define the sensitivity of the analysis at 90\% confidence level as the mean number of pseudo-signal events required to obtain a p-value $\leq 0.5$ in 90\% of the trials. In addition, we define the $3\sigma$ (resp.~$5\sigma$) discovery potential as the mean number of pseudo-signal events required to obtain a p-value $\leq 2.70 \times 10^{-3}$ (resp.~$\leq 5.73 \times 10^{-7}$) in 50\% of the trials. Panel (a) of Fig.~\ref{fig:sensitivity} shows these quantities in terms of the stacked muon-neutrino flux evaluated at\footnote{The shape of the sensitivity and discovery potentials plotted in panel (a) of Fig.~\ref{fig:sensitivity} is a direct consequence of the evaluation of the flux at $E_0 = 10~\mathrm{TeV}$.} $E_0 = 10~\mathrm{TeV}$ as a function of the spectral index $\gamma$. Panel (b) shows the same quantities in terms of the total mean number of pseudo-signal events. We determine this number by integrating the injection spectrum, after convolving it with the detector effective area, over energy and detector lifetime. We find that the sensitivity and discovery potentials are more competitive for harder spectra. This dependence is expected since harder spectra are more easily distinguished from the atmospheric background, \added{which is well-described by an $E^{-3.7}$ power-law spectrum \citep{IceCube_atmospheric}}.

\added{In a more realistic scenario, explored by \citet{Ambrosone_2021}, the spectral index of the power-law spectrum may vary from object to object. For ULIRGs, such a variation could e.g.~be the result of a different relative starburst-versus-AGN contribution to the neutrino flux, since the AGN contribution seems to increase with IR luminosity \citep{Nardini_2010}. The stacked neutrino flux of our selected ULIRGs would therefore be a superposition of these spectra. This effect is also referred to as spectral-index blending \citep{Ambrosone_2021}. In this scenario, the single spectral index $\hat{\gamma}$ fitted in our analysis is therefore slightly biased towards the sources providing most of the neutrino flux at Earth. Our sensitivity in the spectral-index blending scenario will worsen slightly compared to the values given in Fig.~\ref{fig:sensitivity}. However, this reduction of the sensitivity is mitigated by the fact that the the fitted spectral index $\hat{\gamma}$ as well as the sensitivity are mostly driven by the strongest neutrino sources.} 


\section{Results}
\label{sec:results}

\begin{deluxetable}{cc}[h!]
\tablecaption{Resulting flux upper limits at 90\% confidence level of the ULIRG stacking analysis.}
\tablewidth{\textwidth}
\tablehead{ \colhead{Spectral index $\gamma$} & \colhead{$\Phi_{\nu_{\mu} + \bar{\nu}_{\mu}}^{90\%}~ \mathrm{at}~ E_0 = 10~ \mathrm{TeV}$} \\
\colhead{} & \colhead{$\mathrm{10^{-14} \times TeV^{-1}~ cm^{-2}~ s^{-1}}$}
}
\decimals
\startdata
2.0 & 1.23 \\
2.28 & 2.40 \\
2.5 & 3.24 \\
2.87 & 3.62 \\
3.0 & 3.34 \\
\enddata
\tablecomments{The spectral index $\gamma=2.28$ corresponds to the best fit of the diffuse neutrino flux in \citet{IceCube_diffuse_numu_9.5yr}. The spectral index $\gamma=2.87$ corresponds to the best fit in \citet{IceCube_diffuse_HESE_7.5yr}.}
\label{tab:results}
\end{deluxetable}

We report the results of the stacking search for neutrino emission from our representative sample of 75 ULIRGs using 7.5 years of IceCube data, with the total IR flux of each individual source as a stacking weight. The analysis yields a best fit for the number of signal events $\hat{n}_s = 0$, such that the best fit for the spectral index, $\hat{\gamma}$, remains undetermined. The fitted results correspond with a $\mathrm{TS} = 0$, and p-value $=1$. These observations are consistent with the hypothesis that the data is compatible with background. We therefore set upper limits on the stacked muon-neutrino flux of the 75 ULIRGs considered. Since the analysis yields a $\mathrm{TS} = 0$, the upper limits at 90\% confidence level (CL) are set equal to the 90\% sensitivity, shown in Fig.~\ref{fig:sensitivity} for unbroken $E^{-\gamma}$ power-law spectra. Table \ref{tab:results} lists the upper limits, $\Phi_{\nu_{\mu} + \bar{\nu}_{\mu}}^{90\%}$, for some specific values of the spectral index $\gamma$ at the normalization energy $E_0 = 10~\mathrm{TeV}$.

\section{Discussion}
\label{sec:discussion}

\subsection{Limits on the ULIRG Source Population}
\label{sec:limits}

The upper limits $\Phi_{\nu_{\mu} + \bar{\nu}_{\mu}}^{90\%}$ of the ULIRG stacking analysis can be translated into an upper limit on the diffuse muon-neutrino flux originating from all ULIRGs with redshift $z\leq0.13$, computed as $\Phi_{\nu_{\mu} + \bar{\nu}_{\mu}}^{z\leq0.13} = \epsilon_c \Phi_{\nu_{\mu} + \bar{\nu}_{\mu}}^{90\%} / (4\pi~ \mathrm{sr})$. Here, $\epsilon_c = 1.1$ which corrects for the completeness of the ULIRG sample (see Appendix \ref{app:completeness} for more details). This result can be extrapolated to an upper limit on the contribution of the total ULIRG source population to the diffuse neutrino flux. For this extrapolation we assume that all ULIRGs have identical properties of hadronic acceleration and neutrino production over cosmic history. We follow the method presented in \citet{Ahlers_2014} to estimate the neutrino flux of all ULIRGs up to a redshift $z_{\max}$ as
\begin{equation}
    \Phi_{\nu_{\mu} + \bar{\nu}_{\mu}}^{z \leq z_{\max}} = \frac{\xi_{z=z_{\max}}}{\xi_{z=0.13}} \Phi_{\nu_{\mu} + \bar{\nu}_{\mu}}^{z \leq 0.13}.
    \label{eq:flux_population}
\end{equation}
The redshift evolution factor $\xi_z$ effectively integrates the source luminosity function over cosmic history up to a redshift $z$. For an unbroken $E^{-\gamma}$ power-law spectrum, $\xi_z$ becomes energy-independent\added{\footnote{In a spectral-index blending scenario, one has to take into account the spectral-index distribution of the sources when integrating their luminosity function \citep{Ambrosone_2021}.}}. It can then be computed directly given a parameterization $\H(z)$ of the source evolution with redshift \citep{Vereecken_2020}. More details and numerical values of $\xi_z$ are provided in Appendix \ref{app:evolution}.


The ULIRG luminosity function has been a topic of several studies \citep{Kim_1998a,Blain_1999,Genzel_2000,Cowie_2004,Chapman_2005,LeFloch_2005,HopkinsPF_2006,Lonsdale_2006,Caputi_2007,Reddy_2008,Magnelli_2009,Clements_2010,Goto_2011,Magnelli_2011,Casey_2012,Gruppioni_2013}. Qualitatively, it is observed that ULIRGs have a rapidly increasing source evolution up to a redshift $z \sim 1$, followed by a relative flattening of the source evolution at higher redshifts. Unless stated otherwise, in this work we follow \citet{Vereecken_2020} by parameterizing the ULIRG redshift evolution as $\H(z) \propto (1+z)^m$, with $m = 4$ for $z \leq 1$ and $m = 0$ (i.e.~a flat source evolution) for $1<z\leq4$.

\begin{figure*}
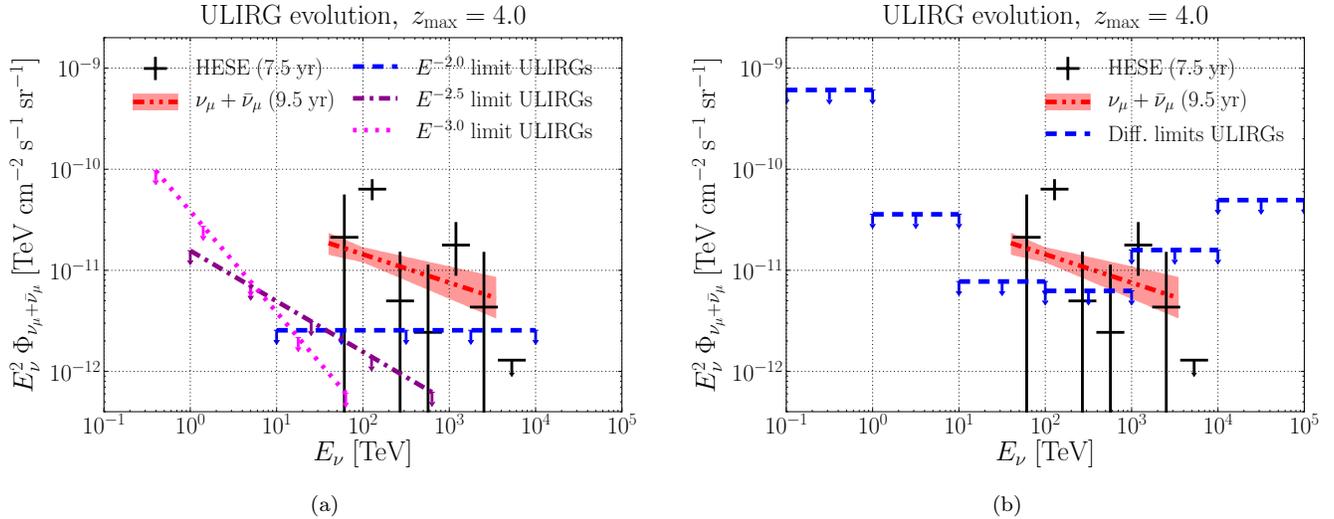

    \centering
    \gridline{\fig{{ULIRG_limit_integral_evolution_ULIRG_zmax_4.0}.pdf}{0.47\textwidth}{(a)}
              \fig{{ULIRG_limit_differential_evolution_ULIRG_zmax_4.0}.pdf}{0.47\textwidth}{(b)}}
    \caption{Panel (a) presents the integral limits at 90\% confidence level on the contribution of the ULIRG population up to a redshift $z_{\max} = 4.0$ to the observed diffuse muon-neutrino flux. These integral limits are shown for unbroken $E^{-2.0}$ (dashed blue line), $E^{-2.5}$ (dash-dotted dark magenta line), and $E^{-3.0}$ (dotted light magenta line) power-law spectra. The integral limits are plotted within their respective 90\% central energy ranges. Panel (b) presents the quasi-differential limits at 90\% confidence level (dashed blue line), found by determining the limit for an $E^{-2.0}$ spectrum in each decade of energy. In each panel, the diffuse neutrino observations are shown in terms of the 7.5 year differential per-flavor measurements using the high-energy starting event (HESE) sample \citep[black data points;][]{IceCube_diffuse_HESE_7.5yr} and the 9.5 year best-fit unbroken power-law spectrum of astrophysical muon neutrinos from the Northern hemisphere \citep[red band;][]{IceCube_diffuse_numu_9.5yr}.}
    \label{fig:population_limits}
\end{figure*}

Using Eq.~(\ref{eq:flux_population}), we are now able to determine the upper limits at 90\% CL~on the diffuse muon-neutrino flux of the ULIRG population up to a redshift $z_{\max} = 4.0$, which is chosen following \citet{Palladino_2019}. The integral limits for unbroken $E^{-2.0}$, $E^{-2.5}$, and $E^{-3.0}$ power-law spectra are shown in panel (a) of Fig.~\ref{fig:population_limits}. Each of these limits is plotted in their respective 90\% central energy region, defined as the energy range in which events contribute to 90\% of the total sensitivity\footnote{The 90\% central energy range of the sensitivity is determined by reducing (resp.~increasing) the maximum (resp.~minimum) allowed energy of injected pseudo-signal events in steps of $\log_{10}(E/\mathrm{GeV}) = 0.2$, and computing the sensitivity in each step. The maximum energy bound (resp.~minimum energy bound) is the value for which the sensitivity flux increases with 5\% compared to the sensitivity flux integrated over all energies.}. We find that the limits under the $E^{-2.0}$ and $E^{-2.5}$ assumptions exclude ULIRGs as the sole contributors to the diffuse neutrino observations up to energies of $\sim$3 PeV and $\sim$600 TeV, respectively. The $E^{-3.0}$ limit mostly constrains the diffuse ULIRG flux at energies below the diffuse observations. Panel (b) of Fig.~\ref{fig:population_limits} shows the quasi-differential limits on the diffuse neutrino flux from ULIRGs. These quasi-differential limits are determined by computing the $E^{-2.0}$ limit in each bin of energy decade between 100 GeV and 10 PeV. We find that the quasi-differential limits \replaced{constrain}{exclude} ULIRGs as the sole contributors to the diffuse neutrino observations in the 10--100 TeV and 0.1--1 PeV bins, respectively.

The above results do not constrain the possible contribution of LIRGs ($10^{11} L_{\odot} \leq L_{\mathrm{IR}} < 10^{12} L_{\odot}$) to the diffuse neutrino flux. Within $z<1$, the IR luminosity density of LIRGs evolves roughly the same with redshift as the IR luminosity density of ULIRGs, although the former is $\sim$10--50 times larger \citep{LeFloch_2005,Caputi_2007,Magnelli_2009,Goto_2011,Casey_2012}. Assuming that the correlation between the total IR and neutrino luminosities holds down to $L_{\mathrm{IR}} \geq 10^{11} L_{\odot}$, the contribution of LIRGs to the neutrino flux is expected to dominate with respect to ULIRGs \added{by a factor $\sim$10--50. Under this assumption, the diffuse neutrino observations would likely provide more stringent constraints than our stacking analysis.} \replaced{However, we note that a possible AGN contribution to the total IR luminosity seems to increase for ULIRGs \citep{Nardini_2010}. This increase in AGN contribution suggests that the hadronic acceleration properties of ULIRGs may differ from those of LIRGs.}{Nevertheless, we note that the AGN contribution to the total IR luminosity seems to be smaller for LIRGs than for ULIRGs \citep{Nardini_2010}. This difference of the AGN contribution suggests that the hadronic acceleration properties of LIRGs may differ from those of ULIRGs.} A dedicated study searching for high-energy neutrinos from LIRGs would provide further insights on the possible acceleration mechanisms within these objects.


\subsection{Comparison with Model Predictions}

\begin{figure*}
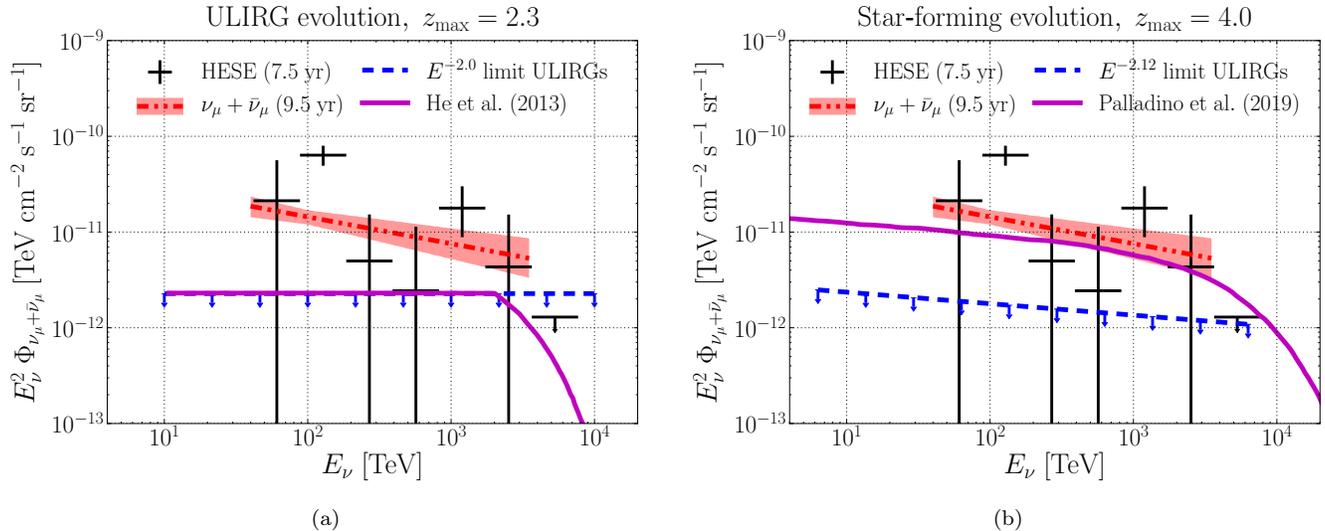

    \centering
    \gridline{\fig{{ULIRG_limit_evolution_ULIRG_gamma_2.0_zmax_2.3_hypernovae}.pdf}{0.47\textwidth}{(a)}
              \fig{{ULIRG_limit_evolution_starforming_gamma_2.12_zmax_4.0_HAGS}.pdf}{0.47\textwidth}{(b)}}
    \caption{Panel (a) shows the prediction by \citet{He_2013} of a diffuse neutrino flux from hypernovae in ULIRGs (full magenta line) and the $E^{-2.0}$ upper limit at 90\% confidence level of this search (dashed blue line). For comparison, an ULIRG source evolution is assumed and integrated up to a redshift $z_{\max} = 2.3$. The model prediction is in tension with the obtained limit. Panel (b) shows the prediction by \citet{Palladino_2019} of a diffuse neutrino flux, normalized to the IceCube observations, from hadronically powered gamma-ray galaxies (HAGS; full magenta line) and the $E^{-2.12}$ upper limit at 90\% confidence level of this search (dashed blue line). For comparison, a source evolution according to the star-formation rate is assumed and integrated up to a redshift $z_{\max} = 4.0$. The obtained limit excludes ULIRGs as the sole HAGS that can explain the diffuse neutrino observations. The diffuse neutrino observations are described in Fig.~\ref{fig:population_limits}.}
    \label{fig:model_limits}
\end{figure*}

First, we compare our results with the cosmic-ray reservoir model developed by \citet{He_2013}. They consider hypernovae as hadronic accelerators which are able to produce $\mathcal{O}(100~\mathrm{PeV})$ protons. \citet{He_2013} suggest that the enhanced star-formation rate in ULIRGs leads to an enhanced hypernova rate. This enhanced hypernova rate results in ULIRG neutrino emission consistent with an $E^{-2.0}$ power-law spectrum, with a cutoff at $\sim$2 PeV. They predict a diffuse neutrino flux originating from ULIRGs up to a redshift $z_{\max} = 2.3$. Hence, we compare the prediction by \citet{He_2013} to our $E^{-2.0}$ upper limit (90\% CL) on the diffuse ULIRG neutrino flux up to a redshift $z_{\max} = 2.3$. This comparison is presented in panel (a) of Fig.~\ref{fig:model_limits}. We find that our upper limit at 90\% CL using 7.5 years of IceCube data is at the level of the predicted flux of \citet{He_2013}. A follow-up study using additional years of data is required to further investigate the validity of this model.

Second, we consider the work of \citet{Palladino_2019}. Their study applies a multimessenger approach to construct a generic cosmic-ray reservoir model of hadronically powered gamma-ray galaxies (HAGS). Candidate HAGS include ULIRGs and starburst galaxies with $L_{\mathrm{IR}} < 10^{12} L_{\odot}$. \citet{Palladino_2019} model the diffuse neutrino and gamma-ray emission of a population of HAGS up to a redshift $z_{\max} = 4.0$ according to a power-law spectrum with an exponential cutoff. This spectrum is fitted to the diffuse IceCube observations given in \citet{IceCube_diffuse_numu_8yr}. They find that spectra with indices $\gamma \leq 2.12$ are able to explain a significant fraction of the diffuse neutrino observations without violating the non-blazar EGB bound above 50 GeV \citep{Fermi_EGB,Lisanti_2016,Zechlin_2016}. In panel (b) of Fig.~\ref{fig:model_limits} we compare the baseline HAGS model with spectral index $\gamma = 2.12$ to our $E^{-2.12}$ upper limit (90\% CL) on the diffuse ULIRG neutrino flux up to a redshift $z_{\max}=4.0$. To determine this limit, we follow \citet{Palladino_2019} and parameterize the ULIRG source evolution according to the star-formation rate, $\H(z) \propto (1+z)^m$, with $m = 3.4$ for $z \leq 1$ and $m = -0.3$ for $1<z\leq4$ \citep{HopkinsAM_2006,Yuksel_2008}. We find that our limits exclude ULIRGs at 90\% CL as the sole HAGS that can be responsible for the diffuse neutrino observations. However, it should be noted that this result does not have any implications for other candidate HAGS, such as starburst galaxies with $L_{\mathrm{IR}} < 10^{12} L_{\odot}$.

Last, we make a comparison with the beam-dump model of \citet{Vereecken_2020}. This model considers Compton-thick AGN, i.e.~AGN obscured by clouds of matter with hydrogen column densities $N_H \gtrsim 1.5 \times 10^{24}~ \mathrm{cm^{-2}}$ \citep{Comastri_2004}, as hadronic accelerators. The neutrino production is modeled through the $pp$-interactions of accelerated hadrons with cloud nuclei. \citet{Vereecken_2020} predict a diffuse neutrino flux from ULIRGs, for which they consider two methods to normalize the proton luminosity. In one approach, it is normalized to the IceCube observations of \citet{IceCube_diffuse_HESE_6yr}. The authors find that ULIRGs can fit the IceCube data without exceeding the $\textit{Fermi}$-LAT non-blazar EGB bound above 50 GeV \citep{Fermi_EGB} for column densities $N_H \gtrsim 5 \times 10^{25}~ \mathrm{cm^{-2}}$. A second approach is found by normalizing the ULIRG proton luminosity to the radio luminosity using the observed ULIRG radio-infrared relation \citep{Sargent_2010}. In this case, the modeling mainly depends on the electron-to-proton luminosity ratio, $f_e$ \citep[see e.g.][]{Merten_2017}.

Here, we fit the results of \citet{Vereecken_2020} for an $E^{-2.0}$ spectrum to our $E^{-2.0}$ upper limit (90\% CL) on the diffuse neutrino flux from ULIRGs up to a redshift $z_{\max} = 4.0$. Subsequently, we use the method presented in \citet{Vereecken_2020} to estimate the ULIRG radio luminosity using the radio-infrared relation of \citet{Sargent_2010}. For this estimation, we take a conservative value $L_{\mathrm{IR}} = 10^{12} L_{\odot}$, and we assume that AGN contribute to 10\% of this infrared luminosity based on the results of \citet{Nardini_2010}. This allows us to set a lower limit on the electron-to-proton luminosity ratio, $f_e \gtrsim 10^{-3}$, by fixing all other parameters in the model of \citet{Vereecken_2020}. It should be noted that among these parameters are quantities with large uncertainties, such as the electron-to-radio luminosity ratio \citep{Tjus_2014}. Hence, the lower limit $f_e \gtrsim 10^{-3}$ should be regarded as an order-of-magnitude estimation. Nevertheless, we point out that this limit on the electron-to-proton luminosity ratio lies within the same order of magnitude as the lower limits provided by \citet{Vereecken_2020} for several obscured flat-spectrum radio AGN that were also studied with IceCube \citep{Maggi_2016,IceCube_obscured_AGN}.

\newpage
\section{Conclusions}
\label{sec:conclusions}

ULIRGs have IR luminosities $L_{\mathrm{IR}} \geq 10^{12} L_{\odot}$, making them the most luminous objects in the IR sky. They are mainly powered by starbursts, possibly combined with an AGN contribution that likely increases with IR luminosity. These starburst and AGN environments are plausible hosts of hadronic accelerators, suggesting that neutrino production occurs in such environments. ULIRGs are therefore candidate sources of high-energy astrophysical neutrinos. Studies by \citet{He_2013}, \citet{Palladino_2019}, and \citet{Vereecken_2020} suggest that the ULIRG population could be responsible for a significant fraction of the diffuse astrophysical neutrino flux observed with the IceCube Neutrino Observatory at the South Pole.

In this work, we presented a stacking search for astrophysical neutrinos from ULIRGs using 7.5 years of IceCube data. A representative sample of 75 ULIRGs with redshifts $z \leq 0.13$ was obtained from three catalogs based on \IRAS~data \citep{Kim_1998a,Sanders_2003,Nardini_2010}. An unbinned maximum likelihood analysis was performed with the IceCube data to search for an excess of an astrophysical signal from ULIRGs above the atmospheric background. No such excess has been found. The analysis yields a $\text{p-value} = 1$, which is consistent with the hypothesis that the data is compatible with background. Hence, we report upper limits (90\% CL) on the neutrino flux originating from these 75 ULIRGs. The stacked flux upper limit for an unbroken $E^{-2.5}$ power-law spectrum is $\Phi_{\nu_\mu + \bar{\nu}_\mu}^{90\%} = 3.24 \times 10^{-14}~ \mathrm{TeV^{-1}~ cm^{-2}~ s^{-1}}~ (E/10~ \mathrm{TeV})^{-2.5}$.

We studied the implication of our null result on the contribution of the ULIRG source population to the IceCube diffuse neutrino observations. The integral limits for unbroken $E^{-2.0}$ and $E^{-2.5}$ power-law spectra exclude ULIRGs as the sole contributors to the diffuse neutrino flux up to energies of $\sim$3 PeV and $\sim$600 TeV, respectively. The quasi-differential limits exclude ULIRGs as the sole contributors to the diffuse neutrino flux in the 10--100 TeV and 0.1--1 PeV energy bins. We remark that these results do not constrain the possible diffuse neutrino contribution of the less luminous but more numerous LIRGs ($10^{11} L_{\odot} \leq L_{\mathrm{IR}} < 10^{12} L_{\odot}$), which have physical properties that are similar to those of ULIRGs \citep[see e.g.][]{Lonsdale_2006}.

Finally, we compared our upper limits at 90\% CL with three models that predict a diffuse neutrino flux from ULIRGs. First, the prediction of \citet{He_2013} is in tension with our $E^{-2.0}$ upper limit, although a follow-up study with additional years of data is required to test the validity of this model. Second, our $E^{-2.12}$ upper limit excludes ULIRGs as the sole population of HAGS, proposed by \citet{Palladino_2019}, that are responsible for the diffuse neutrino observations. We note that this result does not constrain the possible diffuse neutrino contribution from other candidate HAGS, such as starburst galaxies with $L_{\mathrm{IR}} < 10^{12} L_{\odot}$. Third, we report a lower limit on the electron-to-proton luminosity ratio \citep{Merten_2017}, $f_e \gtrsim 10^{-3}$, in the beam-dump model of \citet{Vereecken_2020}. This lower limit was determined by fitting the model to our $E^{-2.0}$ upper limit, while fixing all other model parameters. These parameters include quantities with large uncertainties, such as the electron-to-radio luminosity ratio \citep{Tjus_2014}. Therefore, the lower limit on the electron-to-proton luminosity ratio presented here should be regarded as an order-of-magnitude estimation. A dedicated search for high-energy neutrinos from Compton-thick AGN could provide more insights on the possible neutrino production in such beam-dump scenarios.

\vspace{0.1cm}

\section*{Acknowledgments}
\added{The IceCube collaboration acknowledges the significant contribution to this manuscript from Pablo Correa.} USA {\textendash} U.S. National Science Foundation-Office of Polar Programs,
U.S. National Science Foundation-Physics Division,
U.S. National Science Foundation-EPSCoR,
Wisconsin Alumni Research Foundation,
Center for High Throughput Computing (CHTC) at the University of Wisconsin{\textendash}Madison,
Open Science Grid (OSG),
Extreme Science and Engineering Discovery Environment (XSEDE),
Frontera computing project at the Texas Advanced Computing Center,
U.S. Department of Energy-National Energy Research Scientific Computing Center,
Particle astrophysics research computing center at the University of Maryland,
Institute for Cyber-Enabled Research at Michigan State University,
and Astroparticle physics computational facility at Marquette University;
Belgium {\textendash} Funds for Scientific Research (FRS-FNRS and FWO),
FWO Odysseus and Big Science programmes,
and Belgian Federal Science Policy Office (Belspo);
Germany {\textendash} Bundesministerium f{\"u}r Bildung und Forschung (BMBF),
Deutsche Forschungsgemeinschaft (DFG),
Helmholtz Alliance for Astroparticle Physics (HAP),
Initiative and Networking Fund of the Helmholtz Association,
Deutsches Elektronen Synchrotron (DESY),
and High Performance Computing cluster of the RWTH Aachen;
Sweden {\textendash} Swedish Research Council,
Swedish Polar Research Secretariat,
Swedish National Infrastructure for Computing (SNIC),
and Knut and Alice Wallenberg Foundation;
Australia {\textendash} Australian Research Council;
Canada {\textendash} Natural Sciences and Engineering Research Council of Canada,
Calcul Qu{\'e}bec, Compute Ontario, Canada Foundation for Innovation, WestGrid, and Compute Canada;
Denmark {\textendash} Villum Fonden and Carlsberg Foundation;
New Zealand {\textendash} Marsden Fund;
Japan {\textendash} Japan Society for Promotion of Science (JSPS)
and Institute for Global Prominent Research (IGPR) of Chiba University;
Korea {\textendash} National Research Foundation of Korea (NRF);
Switzerland {\textendash} Swiss National Science Foundation (SNSF);
United Kingdom {\textendash} Department of Physics, University of Oxford.


\appendix

\section{Completeness of the ULIRG Selection}
\label{app:completeness}

\begin{figure}
    \centering
    \includegraphics[width=0.98\linewidth]{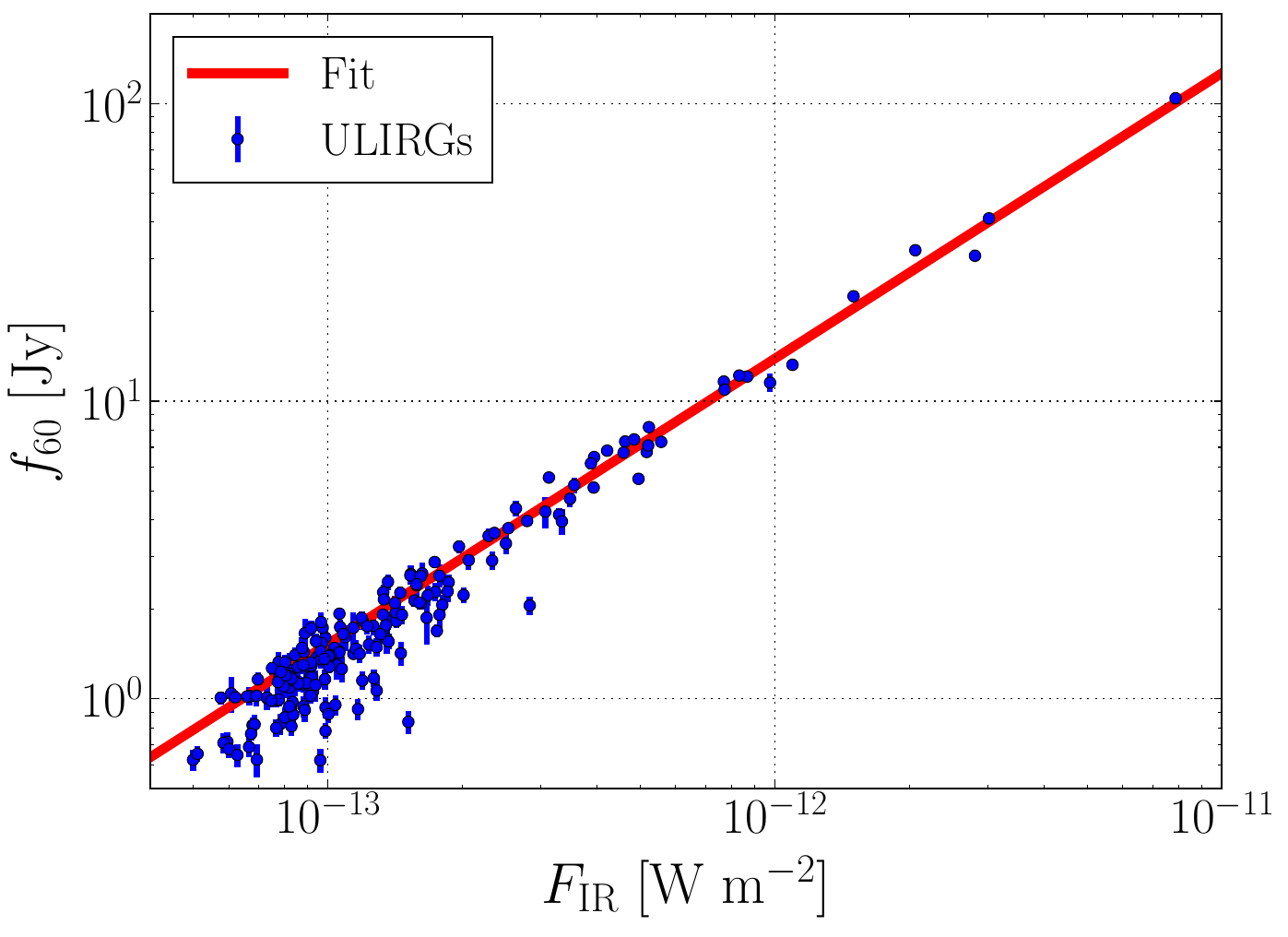}
    \caption{Correlation between the flux density at 60 $\mu$m, $f_{60}$, and the total IR flux between 8--1000 $\mu$m, $F_{\mathrm{IR}}$, for our initial selection of 189 ULIRGs. The log-linear fit through these data points is also shown.}
    \label{fig:flux60_vs_fluxIR}
\end{figure}

A representative sample of the local ULIRG population is found by determining the redshift up to which the least luminous ULIRGs (i.e.~$L_{\mathrm{IR}} = 10^{12} L_{\odot}$) can be observed, given a conservative \IRAS~sensitivity $f_{60} = 1~ \mathrm{Jy}$. For this redshift determination we use the observed correlation between $f_{60}$ and the total IR flux $F_{\mathrm{IR}} = L_{\mathrm{IR}} / (4 \pi d_L^2)$ of our initial ULIRG selection (see Section \ref{sec:ULIRG_selection}), which is shown in Fig.~\ref{fig:flux60_vs_fluxIR}. Here, $d_L$ is the luminosity distance, which is determined from the redshift measurements. Since the $f_{60}$ measurements are optimized separately for the different \IRAS~surveys, these are taken in the following order:
\begin{itemize}
    \item From the RBGS \citep{Sanders_2003} if available;
    \item From the FSC \citep{Moshir_1992} if not available in the RBGS;
    \item From the PSC \citep{Beichman_1988} if not available in the RBGS or FSC.
\end{itemize}
Note that the data from the FSC and PSC are obtained from NED.

To determine the $F_{\mathrm{IR}}$ value corresponding with $f_{60} = 1~ \mathrm{Jy}$, we perform a log-linear fit on the $f_{60}$ and $F_{\mathrm{IR}}$ catalog data of the form
\begin{equation}
    \log_{10} \left( \frac{f_{60}}{\mathrm{Jy}} \right) = a \log_{10} \left( \frac{F_{\mathrm{IR}}}{\mathrm{W~ m^{-2}}} \right) + b,
    \label{eq:completeness}
\end{equation}
with best-fit parameters $a = (95.85 \pm 0.81) \times 10^{-2}$ and $b = 12.645 \pm 0.094$. This fit is shown in Fig.~\ref{fig:flux60_vs_fluxIR}. Subsequently, using Eq.~(\ref{eq:completeness}), the luminosity distance $d_L$ can be determined for which $L_{\mathrm{IR}} = 10^{12} L_{\odot}$ given $f_{60} = 1~ \mathrm{Jy}$. We find that $d_L \sim 700~ \mathrm{Mpc}$, which corresponds to a redshift $z \sim 0.143$. However, the uncertainties on $L_{\mathrm{IR}}$ were not taken into account, since they are not provided by Catalogs 1 \& 2. Therefore, a conservative redshift $z = 0.13$ is adopted up to which we are confident that the ULIRG selection is representative for the local ULIRG population. This conservative redshift cut results in a final selection of 75 ULIRGs with $z \leq 0.13$.

The redshift cut at $z = 0.13$ effectively corresponds with a flux constraint $f_{60} \gtrsim 1~ \mathrm{Jy}$. However, the final sample of 75 ULIRGs likely misses sources with $1~ \mathrm{Jy} \leq f_{60} < 5.24~ \mathrm{Jy}$, called ``1-Jy sources'' hereafter. This lack of sources is due the fact that Catalog 2, although complete, only covers 40\% of the sky. In addition, the required \textit{Spitzer} observations limit the coverage of Catalog 3. The final ULIRG sample contains 37 1-Jy sources from Catalog 2, and 15 1-Jy sources from Catalog 3 which are located in the complementary 60\% of the sky. Hence, we likely miss $\sim$40 1-Jy sources in our final ULIRG selection.

The effect of these missing 1-Jy sources is estimated by computing their contribution to the cumulative stacking weight of the analysis (see Section \ref{sec:analysis_method}). For this estimation, 40 1-Jy sources are simulated evenly over the part of the sky not covered by Catalog 2, where each declination will directly determine the corresponding detector response. Each of these sources is given a theoretical weight equal to the median value of the total IR flux of the 37 ULIRGs taken from Catalog 2. After repeating this simulation $10^4$ times, we find that the median contribution of 40 missing 1-Jy sources to the cumulative stacking weight is roughly 10\% for all spectra considered in this work. Our sample of 75 ULIRGs can therefore still be regarded as a representative sample of the local ULIRG population. We compute the upper limit on the diffuse neutrino flux from all ULIRGs within $z \leq 0.13$ as $\Phi_{\nu_{\mu} + \bar{\nu}_{\mu}}^{z\leq0.13} = \epsilon_c \Phi_{\nu_{\mu} + \bar{\nu}_{\mu}}^{90\%} / (4\pi~ \mathrm{sr})$. Here, $\Phi_{\nu_{\mu} + \bar{\nu}_{\mu}}^{\mathrm{90\%}}$ is the upper limit on the stacked flux from the 75 analyzed ULIRGs (see e.g.~Table \ref{tab:results}), while the completeness correction factor $\epsilon_c = 1.1$ takes into account the missing contribution of $\sim$40 missing 1-Jy sources.

\section{Redshift Evolution Factor}
\label{app:evolution}

A redshift evolution factor $\xi_z$ \citep{Ahlers_2014} was introduced to estimate the muon-neutrino flux corresponding with a certain fraction of the ULIRG source population (see Section \ref{sec:limits}). For an unbroken $E^{-\gamma}$ power-law spectrum, it takes the energy-independent form \citep[e.g.][]{Vereecken_2020}
\begin{equation}
    \xi_z(\gamma) = \int_0^z \frac{\H(z') (1+z')^{-\gamma}}{\sqrt{\Omega_m (1+z')^3 + \Omega_\Lambda}} \d z'.
    \label{eq:redshift_evolution}
\end{equation}
Table \ref{tab:redshift_evolution} lists the values of $\xi_z$ for different combinations of the redshift $z$ up to which Eq.~(\ref{eq:redshift_evolution}) is integrated and the spectral index $\gamma$. These values are given for three parameterizations of the source evolution $\H(z) \propto (1+z)^m $:
\vspace{0.3cm}
\begin{itemize}
    \item ULIRG evolution, where $m = 4$ for $z \leq 1$ and $m = 0$ for $1 < z \leq 4$ \citep{Vereecken_2020}.
    \item Star-forming evolution, where $m = 3.4$ for $z \leq 1$ and $m = -0.3$ for $1 < z \leq 4$ \citep{HopkinsAM_2006,Yuksel_2008}.
    \item Flat evolution, where $m=0$ for all $z$.
\end{itemize}
\vspace{0.3cm}

\onecolumngrid

\vspace{0.4cm}

\begin{deluxetable}{ccDDD}[h!]
\tablecaption{Values of the redshift evolution factor $\xi_z$ for different unbroken $E^{-\gamma}$ power-law spectra and source evolutions.}
\tablewidth{0pt}
\tablehead{
\colhead{Source evolution} & 
\colhead{Spectral index $\gamma$} & \multicolumn2c{$\xi_{z=0.13}$} & 
\multicolumn2c{$\xi_{z=2.3}$} & \multicolumn2c{$\xi_{z=4.0}$}
}
\decimals
\startdata
      & 2.0 & 0.14 & 3.0 & 3.4 \\
ULIRG & 2.5 & 0.14 & 2.2 & 2.5 \\
      & 3.0 & 0.13 & 1.7 & 1.8 \\
\hline
               & 2.0 & 0.14 & 2.2 & 2.4 \\
Star formation & 2.5 & 0.13 & 1.6 & 1.7 \\
               & 3.0 & 0.13 & 1.2 & 1.3 \\
\hline
     & 2.0 & 0.11 & 0.49 & 0.53 \\
Flat & 2.5 & 0.11 & 0.41 & 0.43 \\
     & 3.0 & 0.11 & 0.35 & 0.36 \\
\enddata
\tablecomments{The parameterizations of the considered source evolutions are described in the text.}
\label{tab:redshift_evolution}
\end{deluxetable}

\vspace{-1cm}

\startlongtable
\begin{deluxetable}{cDDDDDDDc}
\tablecaption{ULIRGs used in this IceCube search.}
\tablewidth{0pt}
\tabletypesize{\footnotesize}
\tablehead{
\colhead{NED identification} & 
\multicolumn2c{R.A.} & \multicolumn2c{Dec.} & 
\multicolumn2c{Redshift} & \multicolumn2c{$d_L$} & 
\multicolumn2c{$f_{60}$} & \multicolumn2c{$\sigma_{f_{60}}$} &
\multicolumn2c{$\log_{10} \left( L_{\mathrm{IR}} \right)$} & \colhead{ULIRG catalog} \\ 
\colhead{} & 
\multicolumn2c{deg} & \multicolumn2c{deg} & 
\multicolumn2c{} & \multicolumn2c{Mpc} & 
\multicolumn2c{Jy} & \multicolumn2c{Jy} &
\multicolumn2c{$L_{\odot}$} & \colhead{}
}
\decimalcolnumbers
\startdata
2MASX J00002427-5250313 & 0.10 & -52.84 & 0.125 & 604 & 1.89 & 0.13 & 12.20 & \citet{Nardini_2010} \\
2MASX J00114330-0722073 & 2.93 & -7.37 & 0.118 & 570 & 2.63 & 0.18 & 12.19 & \citet{Kim_1998a} \\
2MASX J00212652-0839261 & 5.36 & -8.66 & 0.128 & 622 & 2.59 & 0.23 & 12.33 & \citet{Kim_1998a} \\
2MASX J00220698-7409418 & 5.53 & -74.16 & 0.096 & 457 & 4.16 & 0.21 & 12.33 & \citet{Nardini_2010} \\
2MASX J00480675-2848187 & 12.03 & -28.81 & 0.110 & 526 & 2.60 & 0.18 & 12.12 & \citet{Kim_1998a} \\
GALEXASC J005040.35-270440.6 & 12.67 & -27.08 & 0.129 & 626 & 1.13 & 0.14 & 12.00 & \citet{Kim_1998a} \\
GALEXASC J010250.01-222157.4 & 15.71 & -22.37 & 0.118 & 567 & 2.29 & 0.16 & 12.24 & \citet{Kim_1998a} \\
2MASX J01190760-0829095 & 19.78 & -8.49 & 0.118 & 569 & 1.74 & 0.10 & 12.03 & \citet{Kim_1998a} \\
2MASX J01405591-4602533 & 25.23 & -46.05 & 0.090 & 426 & 2.91 & 0.20 & 12.12 & \citet{Nardini_2010} \\
2MASX J02042730-2049413 & 31.11 & -20.83 & 0.116 & 557 & 1.45 & 0.12 & 12.01 & \citet{Kim_1998a} \\
2MASX J03274981+1616594 & 51.96 & 16.28 & 0.129 & 625 & 1.38 & 0.08 & 12.06 & \citet{Kim_1998a} \\
2MASX J04121945-2830252 & 63.08 & -28.51 & 0.117 & 565 & 1.82 & 0.09 & 12.15 & \citet{Kim_1998a} \\
2MASX J04124420-5109402 & 63.18 & -51.16 & 0.125 & 602 & 2.09 & 0.10 & 12.26 & \citet{Nardini_2010} \\
2MASX J05210136-2521450 & 80.26 & -25.36 & 0.043 & 194 & 13.25 & 0.03 & 12.11 & \citet{Sanders_2003} \\
2MASX J05583717-7716393 & 89.65 & -77.28 & 0.117 & 562 & 1.42 & 0.06 & 12.05 & \citet{Nardini_2010} \\
2MASX J06025406-7103104 & 90.73 & -71.05 & 0.079 & 373 & 5.13 & 0.15 & 12.23 & \citet{Nardini_2010} \\
2MASX J06210118-6317238 & 95.26 & -63.29 & 0.092 & 437 & 3.96 & 0.12 & 12.22 & \citet{Nardini_2010} \\
2MASX J07273754-0254540 & 111.91 & -2.92 & 0.088 & 413 & 6.49 & 0.03 & 12.32 & \citet{Sanders_2003} \\
2MASX J08380365+5055090 & 129.52 & 50.92 & 0.097 & 459 & 2.14 & 0.11 & 12.01 & \citet{Nardini_2010} \\
IRAS 08572+3915 & 135.11 & 39.07 & 0.058 & 270 & 7.30 & 0.03 & 12.10 & \citet{Sanders_2003} \\
2MASX J09041268-3627007 & 136.05 & -36.45 & 0.060 & 276 & 11.64 & 0.06 & 12.26 & \citet{Sanders_2003} \\
2MASX J09063400+0451271 & 136.64 & 4.86 & 0.125 & 605 & 1.48 & 0.09 & 12.07 & \citet{Kim_1998a} \\
2MASX J09133888-1019196 & 138.41 & -10.32 & 0.054 & 249 & 6.75 & 0.04 & 12.00 & \citet{Sanders_2003} \\
UGC 05101 & 143.96 & 61.35 & 0.039 & 179 & 11.54 & 0.81 & 11.99 & \citet{Nardini_2010} \\
2MASX J09452133+1737533 & 146.34 & 17.63 & 0.128 & 621 & 0.89 & 0.06 & 12.08 & \citet{Nardini_2010} \\
GALEXASC J095634.42+084306.0 & 149.14 & 8.72 & 0.129 & 624 & 1.44 & 0.10 & 12.03 & \citet{Kim_1998a} \\
IRAS 10190+1322 & 155.43 & 13.12 & 0.077 & 358 & 3.33 & 0.27 & 12.00 & \citet{Kim_1998a} \\
2MASX J10522356+4408474 & 163.10 & 44.15 & 0.092 & 435 & 3.53 & 0.21 & 12.13 & \citet{Kim_1998a} \\
2MASX J10591815+2432343 & 164.83 & 24.54 & 0.043 & 197 & 12.10 & 0.03 & 12.02 & \citet{Sanders_2003} \\
LCRS B110930.3-023804 & 168.01 & -2.91 & 0.107 & 509 & 3.25 & 0.16 & 12.20 & \citet{Kim_1998a} \\
2MASX J11531422+1314276 & 178.31 & 13.24 & 0.127 & 616 & 2.58 & 0.15 & 12.28 & \citet{Kim_1998a} \\
IRAS 12071-0444 & 182.44 & -5.02 & 0.128 & 621 & 2.46 & 0.15 & 12.35 & \citet{Kim_1998a} \\
IRAS 12112+0305 & 183.44 & 2.81 & 0.073 & 342 & 8.18 & 0.03 & 12.28 & \citet{Sanders_2003} \\
Mrk 0231 & 194.06 & 56.87 & 0.042 & 193 & 30.80 & 0.04 & 12.51 & \citet{Sanders_2003} \\
WKK 2031 & 198.78 & -55.16 & 0.031 & 139 & 41.11 & 0.07 & 12.26 & \citet{Sanders_2003} \\
IRAS 13335-2612 & 204.09 & -26.46 & 0.125 & 604 & 1.40 & 0.11 & 12.06 & \citet{Kim_1998a} \\
Mrk 0273 & 206.18 & 55.89 & 0.038 & 172 & 22.51 & 0.04 & 12.14 & \citet{Sanders_2003} \\
4C +12.50 & 206.89 & 12.29 & 0.122 & 587 & 1.92 & 0.21 & 12.28 & \citet{Kim_1998a} \\
IRAS 13454-2956 & 207.08 & -30.20 & 0.129 & 625 & 2.16 & 0.11 & 12.21 & \citet{Kim_1998a} \\
2MASX J13561001+2905355 & 209.04 & 29.09 & 0.108 & 518 & 1.83 & 0.13 & 12.00 & \citet{Kim_1998a} \\
2MASXJ 14081899+2904474 & 212.08 & 29.08 & 0.117 & 561 & 1.61 & 0.16 & 12.03 & \citet{Kim_1998a} \\
IRAS 14348-1447 & 219.41 & -15.01 & 0.083 & 390 & 6.82 & 0.04 & 12.30 & \citet{Sanders_2003} \\
2MASX J14405901-3704322 & 220.25 & -37.08 & 0.068 & 315 & 6.72 & 0.04 & 12.15 & \citet{Sanders_2003} \\
2MASX J14410437+5320088 & 220.27 & 53.34 & 0.105 & 498 & 1.95 & 0.08 & 12.04 & \citet{Kim_1998a} \\
2MASX J15155520-2009172 & 228.98 & -20.15 & 0.109 & 520 & 1.92 & 0.13 & 12.09 & \citet{Kim_1998a} \\
SDSS J152238.10+333135.8 & 230.66 & 33.53 & 0.124 & 601 & 1.77 & 0.09 & 12.18 & \citet{Kim_1998a} \\
IRAS 15250+3609 & 231.75 & 35.98 & 0.055 & 254 & 7.10 & 0.04 & 12.02 & \citet{Sanders_2003} \\
Arp 220 & 233.74 & 23.50 & 0.018 & 81 & 104.09 & 0.11 & 12.21 & \citet{Sanders_2003} \\
IRAS 15462-0450 & 237.24 & -4.99 & 0.100 & 474 & 2.92 & 0.20 & 12.16 & \citet{Kim_1998a} \\
2MASX J16491420+3425096 & 252.31 & 34.42 & 0.111 & 534 & 2.27 & 0.11 & 12.11 & \citet{Kim_1998a} \\
SBS 1648+547 & 252.45 & 54.71 & 0.104 & 494 & 2.88 & 0.12 & 12.12 & \citet{Kim_1998a} \\
2MASX J17034196+5813443 & 255.92 & 58.23 & 0.106 & 506 & 2.43 & 0.15 & 12.10 & \citet{Kim_1998a} \\
2MASX J17232194-0017009 & 260.84 & -0.28 & 0.043 & 196 & 32.13 & 0.06 & 12.39 & \citet{Sanders_2003} \\
2MASX J18383543+3552197 & 279.65 & 35.87 & 0.116 & 558 & 2.23 & 0.13 & 12.29 & \citet{Nardini_2010} \\
IRAS 18588+3517 & 285.17 & 35.36 & 0.107 & 510 & 1.47 & 0.10 & 11.97 & \citet{Nardini_2010} \\
Superantennae & 292.84 & -72.66 & 0.062 & 286 & 5.48 & 0.22 & 12.10 & \citet{Nardini_2010} \\
2MASX J19322229-0400010 & 293.09 & -4.00 & 0.086 & 404 & 7.32 & 0.11 & 12.37 & \citet{Sanders_2003} \\
IRAS 19458+0944 & 297.07 & 9.87 & 0.100 & 475 & 3.95 & 0.40 & 12.37 & \citet{Nardini_2010} \\
IRAS 19542+1110 & 299.15 & 11.32 & 0.065 & 301 & 6.18 & 0.04 & 12.04 & \citet{Sanders_2003} \\
2MASX J20112386-0259503 & 302.85 & -3.00 & 0.106 & 504 & 4.70 & 0.28 & 12.44 & \citet{Nardini_2010} \\
2MASX J20132950-4147354 & 303.37 & -41.79 & 0.130 & 628 & 5.23 & 0.31 & 12.64 & \citet{Nardini_2010} \\
IRAS 20414-1651 & 311.08 & -16.67 & 0.087 & 410 & 4.36 & 0.26 & 12.14 & \citet{Kim_1998a} \\
ESO 286-IG 019 & 314.61 & -42.65 & 0.043 & 196 & 12.19 & 0.03 & 12.00 & \citet{Sanders_2003} \\
IRAS 21208-0519 & 320.87 & -5.12 & 0.130 & 630 & 1.17 & 0.07 & 12.01 & \citet{Kim_1998a} \\
$\mathrm{ \left[HB89\right] }$ 2121-179 & 321.17 & -17.75 & 0.112 & 536 & 1.07 & 0.09 & 12.06 & \citet{Kim_1998a} \\
2MASX J21354580-2332359 & 323.94 & -23.54 & 0.125 & 604 & 1.65 & 0.08 & 12.09 & \citet{Kim_1998a} \\
2MASX J22074966+3039393 & 331.96 & 30.66 & 0.127 & 614 & 1.87 & 0.36 & 12.29 & \citet{Nardini_2010} \\
IRAS 22491-1808 & 342.96 & -17.87 & 0.078 & 364 & 5.54 & 0.04 & 12.11 & \citet{Sanders_2003} \\
2MASX J23042114+3421477 & 346.09 & 34.36 & 0.108 & 516 & 1.42 & 0.10 & 11.99 & \citet{Nardini_2010} \\
ESO 148-IG 002 & 348.94 & -59.05 & 0.045 & 204 & 10.94 & 0.04 & 12.00 & \citet{Sanders_2003} \\
2MASX J23254938+2834208 & 351.46 & 28.57 & 0.114 & 547 & 1.26 & 0.13 & 12.00 & \citet{Kim_1998a} \\
2MASX J23255611+1002500 & 351.48 & 10.05 & 0.128 & 619 & 1.56 & 0.09 & 12.05 & \citet{Kim_1998a} \\
2MASX J23260362-6910185 & 351.52 & -69.17 & 0.107 & 509 & 3.74 & 0.15 & 12.31 & \citet{Nardini_2010} \\
2MASX J23351192+2930000 & 353.80 & 29.50 & 0.107 & 511 & 2.10 & 0.13 & 12.06 & \citet{Kim_1998a} \\
2MASX J23390127+3621087 & 354.76 & 36.35 & 0.064 & 299 & 7.44 & 0.05 & 12.13 & \citet{Sanders_2003} \\
\enddata
\tablecomments{Column (1): Primary name identification in the NASA/IPAC Extragalactic Database (NED). Columns (2)--(4): Equatorial J2000 coordinates and redshift taken from NED. Column (5): Luminosity distance determined from the redshift measurements using the cosmological parameters in \citet{Planck_2015}. Columns (6)-(7): Infrared flux at 60 $\mu$m and its uncertainty. Taken from the \IRAS~Revised Bright Galaxy Sample \citep[RBGS;][]{Sanders_2003} if available; taken from the \IRAS~Faint Source Catalog \citep[FSC;][]{Moshir_1992} if not available in the RBGS; taken from the \IRAS~Point Source Catalog \citep[PSC;][]{Beichman_1988} if not available in the RBGS or FSC. Data from the FSC and PSC is obtained from NED. Columns (8)--(9): Total infrared luminosity between 8--1000 $\mu$m and the ULIRG catalog from which this value is taken.}
\label{tab:ulirgs}
\end{deluxetable}

\bibliography{bibliography}{}
\bibliographystyle{aasjournal}

\listofchanges
\end{document}